# Auxiliary matrix formalism for interaction representation transformations, optimal control and spin relaxation theories

D.L. Goodwin, Ilya Kuprov[*]

*School of Chemistry, University of Southampton,
Highfield Campus, Southampton, SO17 1BJ, UK.*

[*]Corresponding author (i.kuprov@soton.ac.uk )

**Abstract**

Auxiliary matrix exponential method is used to derive simple and numerically efficient general expressions for the following, historically rather cumbersome and hard to compute, theoretical methods: (1) average Hamiltonian theory following interaction representation transformations; (2) Bloch-Redfield-Wangsness theory of nuclear and electron relaxation; (3) gradient ascent pulse engineering version of quantum optimal control theory. In the context of spin dynamics, the auxiliary matrix exponential method is more efficient than methods based on matrix factorizations and also exhibits more favourable complexity scaling with the dimension of the Hamiltonian matrix.

**Introduction**

Among the many complicated functions encountered in magnetic resonance simulation context, chained exponential integrals involving square matrices $\mathbf{A}_k$ and $\mathbf{B}_k$ occur particularly often:

$$\int_0^t dt_1 \int_0^{t_1} dt_2 \ldots \int_0^{t_{n-2}} dt_{n-1} \left\{ e^{\mathbf{A}_1(t-t_1)} \mathbf{B}_1 e^{\mathbf{A}_2(t_1-t_2)} \mathbf{B}_2 \ldots \mathbf{B}_{n-1} e^{\mathbf{A}_n t_{n-1}} \right\}. \quad (1)$$

Examples include perturbative relaxation theories[1-3], reaction yield expressions in radical pair dynamics[4-6], average Hamiltonian theory[7], fidelity functional derivatives in optimal control theory[8,9] and pulsed field gradient propagators in nuclear magnetic resonance[10]. Their common feature is the complexity of evaluation: expensive matrix factorizations[*] are usually required[3,11,12]. This makes the application of the associated theories difficult when matrix dimension exceeds $10^3$, *i.e.* for ten spins or more.

Consider the example of spin relaxation theory. The currently used techniques for the evaluation of Redfield's integral, which involves a static Hamiltonian $\mathbf{H}$, a rotational correlation function $G(\tau)$ and an irreducible spherical component $\mathbf{Q}$ of the stochastic Hamiltonian, either involve the diagonalization of $\mathbf{H}$ followed by the evaluation of a large number of Fourier transforms[1,3]:

$$\left[ \int_0^\infty G(\tau) e^{-i\mathbf{H}\tau} \mathbf{Q}^\dagger e^{i\mathbf{H}\tau} d\tau \right]_{nk} = \left[ \int_0^\infty G(\tau) \left[ \mathbf{V} e^{-i\mathbf{D}\tau} \mathbf{V}^\dagger \right] \mathbf{Q}^\dagger \left[ \mathbf{V} e^{i\mathbf{D}\tau} \mathbf{V}^\dagger \right] d\tau \right]_{nk} =$$

$$= \left[ \int_0^\infty G(\tau) \mathbf{V} e^{-i\mathbf{D}\tau} \left[ \mathbf{V}^\dagger \mathbf{Q}^\dagger \mathbf{V} \right] e^{i\mathbf{D}\tau} \mathbf{V}^\dagger d\tau \right]_{nk} = \sum_{rs} \int_0^\infty G(\tau) V_{nr} e^{-iD_{rr}\tau} \left[ \mathbf{V}^\dagger \mathbf{Q}^\dagger \mathbf{V} \right]_{rs} e^{iD_{ss}\tau} V_{ks}^* d\tau = \quad (2)$$

$$= \sum_{rs} V_{nr} \left[ \mathbf{V}^\dagger \mathbf{Q}^\dagger \mathbf{V} \right]_{rs} V_{ks}^* \int_0^\infty G(\tau) e^{-i(D_{rr}-D_{ss})\tau} d\tau; \qquad \mathbf{H} = \mathbf{V}\mathbf{D}\mathbf{V}^\dagger$$

or a matrix-valued numerical quadrature with a large number of time steps[13]. The latter method scales better because matrix sparsity is preserved at every stage and diagonalization is avoided, but the evaluation is still difficult. Such situations are ubiquitous in magnetic resonance and the integrals in Equation (1) are the bottleneck in many practically important cases. Ideally, their evaluation should be an elementary function that does not involve either expensive matrix operations or numerical quadrature grids.

In this communication we propose a solution to this problem, based on the observation that matrix exponentiation, when used judiciously, does not require factorizations and preserves spin operator sparsity[13,14] and on the auxiliary matrix technique[15-17] for the evaluation of the integrals given in

---

[*] An "expensive" matrix operation in this context is defined as requiring computational or storage resources that grow faster than quadratically with the dimension of the matrix; a "cheap" matrix operation is defined as the one with approximately linear resource requirements.

Equation (1). The result is a simplification, automation and numerical acceleration of some of the oldest and most useful magnetic resonance theories, particularly for systems involving large numbers of interacting spins. This work is motivated in particular by the practical needs arising during the development of *Spinach*[14], which is a large-scale spin dynamics simulation library that cannot, as a matter of general policy, rely on expensive matrix operations or expect the user to pre-process the Hamiltonian by hand. The methods described below were found, in our practical simulation work, to be user- and developer-friendly as well as numerically efficient.

**Exponentials of auxiliary matrices**

A method for computing some of the integrals of the general type shown in Equation (1) was proposed by Van Loan in 1978[15]. He noted that the integrals in question are solutions to linear block matrix differential equations and suggested that block matrix exponentials are used to compute them. In the simplest case of a single integral:

$$\exp\left[\begin{pmatrix} \mathbf{A} & \mathbf{B} \\ \mathbf{0} & \mathbf{C} \end{pmatrix} t\right] = \begin{pmatrix} e^{\mathbf{A}t} & e^{\mathbf{A}t}\int_0^t e^{-\mathbf{A}t_1}\mathbf{B}e^{\mathbf{C}t_1}dt_1 \\ \mathbf{0} & e^{\mathbf{C}t} \end{pmatrix}. \tag{3}$$

Van Loan's method was subsequently refined by Carbonell *et al.*, who derived a convenient expression using the exponential of a block-bidiagonal auxiliary matrix[17]:

$$\mathbf{M} = \begin{pmatrix} \mathbf{A}_{11} & \mathbf{A}_{12} & 0 & 0 & 0 \\ 0 & \mathbf{A}_{22} & \mathbf{A}_{23} & 0 & 0 \\ 0 & 0 & \mathbf{A}_{33} & \ddots & 0 \\ 0 & 0 & 0 & \ddots & \mathbf{A}_{k-1,k} \\ 0 & 0 & 0 & 0 & \mathbf{A}_{kk} \end{pmatrix}, \quad \exp(\mathbf{M}t) = \begin{pmatrix} \mathbf{B}_{11} & \mathbf{B}_{12} & \mathbf{B}_{13} & \cdots & \mathbf{B}_{1k} \\ 0 & \mathbf{B}_{22} & \mathbf{B}_{23} & \cdots & \mathbf{B}_{2k} \\ 0 & 0 & \mathbf{B}_{33} & \ddots & \vdots \\ 0 & 0 & 0 & \ddots & \mathbf{B}_{k-1,k} \\ 0 & 0 & 0 & 0 & \mathbf{B}_{kk} \end{pmatrix} \tag{4}$$

The integrals in question populate the rows of the resulting block matrix, *e.g.*:

$$\mathbf{B}_{1k} = \int_0^t dt_1 \int_0^{t_1} dt_2 \ldots \int_0^{t_{k-2}} dt_{k-1} \left\{ e^{\mathbf{A}_{11}(t-t_1)} \mathbf{A}_{12} e^{\mathbf{A}_{22}(t_1-t_2)} \mathbf{A}_{23} \ldots \mathbf{A}_{k-1,k} e^{\mathbf{A}_{kk}t_{k-1}} \right\} \tag{5}$$

Equations (4) and (5) are particularly appealing in the context of spin dynamics – spin Hamiltonians are guaranteed to be sparse in the Pauli basis[18-20] and their exponential propagators are also sparse when $\|\mathbf{H}\Delta t\| < 1$, if care is taken to eliminate insignificant (defined as "too small to affect the dynamics on the time scale of the simulation") elements after each matrix multiplication in the scaled and squared Taylor series procedure[14]:

$$e^{\mathbf{M}t} = \sum_{k=0}^{\infty} \frac{(\mathbf{M}t)^k}{k!}; \qquad e^{\mathbf{M}t} = \left(\left(\left(e^{\mathbf{M}t/2^n}\right)^2\right)^2 \ldots\right)^2 \tag{6}$$

Out of the multitude of "dubious ways"[21,22] of computing matrix exponentials, the Taylor series method with scaling and squaring is recommended here because it is compatible with dissipative dynamics (Chebyshev series diverge with non-Hermitian matrices[23]), only involves matrix multiplications (Padé method requires a costly and perilous inverse[15]), uses minimal memory resources (Newton polynomials are more expensive[24]) and only requires approximate scaling (Newton and Chebyshev methods are less forgiving[23,24]). For a publicly available simulation package that must preserve sparsity (matrix dimensions for large NMR systems are in the millions[25]) and run reliably in a large variety of settings[14,19], these considerations are decisive.

We demonstrate below that Equations (4)-(6) remove the problems associated with the calculation of nested exponential integrals from several widely used magnetic resonance simulation methods. The rest of this paper goes through those we could identify and reformulates them to use the auxiliary matrix formalism.

**Rotating frame transformations**

A common first stage of quantum dynamics simulations is the interaction representation transformation wherein the Hamiltonian matrix is split into the "large and simple" part $\mathbf{H}_0$ and the "small and complicated" part $\mathbf{H}_1$, and the following unitary transformation is applied to the Liouville - von Neumann equation of motion[11,12]:

$$\boldsymbol{\sigma}(t) = e^{i\mathbf{H}_0 t}\boldsymbol{\rho}(t)e^{-i\mathbf{H}_0 t} \qquad \mathbf{H}_1^R(t) = e^{i\mathbf{H}_0 t}\mathbf{H}_1 e^{-i\mathbf{H}_0 t}$$
$$\frac{d}{dt}\boldsymbol{\rho}(t) = -i\left[\mathbf{H}_0 + \mathbf{H}_1, \boldsymbol{\rho}(t)\right] \quad \Rightarrow \quad \frac{d}{dt}\boldsymbol{\sigma}(t) = -i\left[\mathbf{H}_1^R(t), \boldsymbol{\sigma}(t)\right] \qquad (7)$$

where $\boldsymbol{\rho}(t)$ is the density operator. An important next step is to make $\mathbf{H}_1^R(t)$ effectively time-independent by averaging its exponential propagator over the period of $\exp(-i\mathbf{H}_0 t)$. This is achieved by combining the solution of the laboratory frame Liouville - von Neumann equation

$$\boldsymbol{\rho}(t) = e^{-i(\mathbf{H}_0+\mathbf{H}_1)t}\boldsymbol{\rho}(0)e^{i(\mathbf{H}_0+\mathbf{H}_1)t} \qquad (8)$$

with the definition of the interaction representation (*aka* "rotating frame")

$$\boldsymbol{\sigma}(t) = e^{i\mathbf{H}_0 t}\boldsymbol{\rho}(t)e^{-i\mathbf{H}_0 t} \qquad (9)$$

and observing that $\boldsymbol{\sigma}(0) = \boldsymbol{\rho}(0)$:

$$\boldsymbol{\sigma}(t) = e^{i\mathbf{H}_0 t}e^{-i(\mathbf{H}_0+\mathbf{H}_1)t}\boldsymbol{\sigma}(0)e^{i(\mathbf{H}_0+\mathbf{H}_1)t}e^{-i\mathbf{H}_0 t} = \mathbf{U}(t)\boldsymbol{\sigma}(0)\mathbf{U}^\dagger(t) \qquad (10)$$

The effective propagator in the rotating frame is therefore

$$\mathbf{U}(t) = e^{i\mathbf{H}_0 t}e^{-i(\mathbf{H}_0+\mathbf{H}_1)t} \qquad (11)$$

and the effective rotating frame Hamiltonian over a time interval $[0,T]$ is the principal value of the logarithm of $\mathbf{U}(T)$

$$\bar{\mathbf{H}} = \frac{i}{T}\ln_{(P)}\left[e^{i\mathbf{H}_0 T}e^{-i(\mathbf{H}_0+\mathbf{H}_1)T}\right] \tag{12}$$

If $T$ is now chosen to be the *period* of $\exp(-i\mathbf{H}_0 t)$, the propagator $\exp(-i\mathbf{H}_0 T)$ becomes equal to the unit matrix and Equation (12) acquires a form similar to that seen in the theory of generalized cumulant expansions[2,26]:

$$\bar{\mathbf{H}} = \frac{i}{T}\ln_{(P)}\left[e^{-i(\mathbf{H}_0+\mathbf{H}_1)T}\right] \tag{13}$$

in which the reader should resist the temptation to cancel the exponential and the logarithm – with the above mentioned period condition on $T$, the principal value of the logarithm is not equal to the matrix that has gone into the exponential[27]. Equation (13) is a very compact formulation of the exact average Hamiltonian over the period of the rotating frame. It also provides a useful generating function for the perturbative expansion that we are about to derive.

Under the typical interaction representation assumptions, $\|\mathbf{H}_1\| < \|\mathbf{H}_0\|$, and therefore the $\mathbf{H}_1$ term under the exponential in Equation (13) is a correction to the $\mathbf{H}_0$ term. The corresponding Taylor series with respect to the $\mathbf{H}_1$ direction step length parameter $\alpha$ is:

$$\bar{\mathbf{H}}(\alpha) = \frac{i}{T}\ln\left[e^{-i(\mathbf{H}_0+\alpha\mathbf{H}_1)T}\right] = \frac{i}{T}\sum_{n=1}^{\infty}\frac{\partial^n}{\partial\alpha^n}\left(\ln\left[e^{-i(\mathbf{H}_0+\alpha\mathbf{H}_1)T}\right]\right)\bigg|_{\alpha=0}\frac{\alpha^n}{n!} \tag{14}$$

The logarithm disappears after the first differentiation, leaving simple expressions for the perturbative corrections to the effective Hamiltonian in which nested commutators do not occur, the number of matrix terms is linear with respect to the approximation order $n$ and further terms may be obtained by repeated application of the product rule:

$$\begin{aligned}\bar{\mathbf{H}}^{(1)} &= \frac{i}{T}\mathbf{D}_0^\dagger\mathbf{D}_1 \\ \bar{\mathbf{H}}^{(2)} &= \frac{i}{2T}\left(\mathbf{D}_1^\dagger\mathbf{D}_1+\mathbf{D}_0^\dagger\mathbf{D}_2\right) \qquad\qquad \mathbf{D}_k = \frac{\partial^k e^{-i(\mathbf{H}_0+\alpha\mathbf{H}_1)T}}{\partial\alpha^k}\bigg|_{\alpha=0} \\ \bar{\mathbf{H}}^{(3)} &= \frac{i}{6T}\left(\mathbf{D}_2^\dagger\mathbf{D}_1+2\mathbf{D}_1^\dagger\mathbf{D}_2+\mathbf{D}_0^\dagger\mathbf{D}_3\right)\end{aligned} \tag{15}$$

with the general expression also obtainable using binomial coefficients:

$$\bar{\mathbf{H}} = \frac{i}{T}\sum_{n=1}^{\infty}\frac{1}{n}\sum_{k=1}^{n}\frac{\mathbf{D}_{n-k}^\dagger\mathbf{D}_k}{(k-1)!(n-k)!} \ . \tag{16}$$

The simplicity of Equation (16) stands in sharp contrast with the very large expressions produced by Magnus expansions. The derivatives $\mathbf{D}_k$ of $\exp\left[-i(\mathbf{H}_0 + \alpha\mathbf{H}_1)T\right]$ with respect to $\alpha$ at $\alpha = 0$ are known[16] to have the form that matches the auxiliary matrix integrals in Equation (5):

$$\mathbf{D}_k(T) = k e^{\mathbf{A}T} \int_0^T e^{-\mathbf{A}t} \mathbf{B} \mathbf{D}_{k-1}(t) dt, \quad (17)$$

$$\mathbf{D}_0(t) = e^{\mathbf{A}t}, \qquad \mathbf{A} = -i\mathbf{H}_0, \qquad \mathbf{B} = -i\mathbf{H}_1$$

This makes them easy to compute by constructing and exponentiating a very sparse block-bidiagonal matrix $\mathbf{M}$ prescribed by Equation (4):

$$\mathbf{M} = \begin{pmatrix} \mathbf{A} & \mathbf{B} & 0 & 0 & 0 \\ 0 & \mathbf{A} & \mathbf{B} & 0 & 0 \\ 0 & 0 & \mathbf{A} & \ddots & 0 \\ 0 & 0 & 0 & \ddots & \mathbf{B} \\ 0 & 0 & 0 & 0 & \mathbf{A} \end{pmatrix} \quad \exp(t\mathbf{M}) = \begin{pmatrix} \mathbf{D}_0 & \mathbf{D}_1/1! & \mathbf{D}_2/2! & \cdots & \mathbf{D}_k/k! \\ 0 & \mathbf{D}_0 & \mathbf{D}_1/1! & \cdots & \mathbf{D}_{k-1}/(k-1)! \\ 0 & 0 & \mathbf{D}_0 & \ddots & \vdots \\ 0 & 0 & 0 & \ddots & \mathbf{D}_1/1! \\ 0 & 0 & 0 & 0 & \mathbf{D}_0 \end{pmatrix} \quad (18)$$

and extracting the first block row from the result. Because $T$ is a period of $\exp(-i\mathbf{H}_0 t)$ and $\|\mathbf{H}_1\| < \|\mathbf{H}_0\|$, the 2-norm of $t\mathbf{M}$ is approximately $2\pi$, meaning that the exponential of $t\mathbf{M}$ is also sparse[20] if due care is taken to eliminate inconsequentially small elements from the non-zero index after each multiplication in Equation (6). For the same reason, the $\mathbf{D}_0$ terms in Equations (15)-(18) are actually unit matrices. The number of blocks in $\mathbf{M}$ grows linearly and the effort of exponentiating it approximately quadratically (Figure 1A) with the rotating frame correction order $n$.

The primary advantage of this path through the average Hamiltonian theory[7] is the simplicity of implementation and the possibility of automation – perturbative corrections to the rotating frame transformation used to be an arduous manual process that had to be endured afresh for each new class of magnetic resonance problems. The procedure described above is also an improvement in the sense that its numerical implementations require no knowledge of the content of $\mathbf{H}_0$ and $\mathbf{H}_1$ – it may be incorporated into highly general simulation codes without the logistical overhead of storing every significant magnetic resonance assumption set and rotating frame layout.

An example illustrating the practical convenience and efficiency of Equation (16) is given in Figure 1A – for the spin system of methylaziridine (state space dimension 9,889 using IK-2(4) basis[25] in Liouville space[28]), the rotating frame transformation with respect to the Zeeman Hamiltonian of the $^{14}$N nucleus is computed in seconds and the scaling of the wall clock time with respect to the approximation order is quadratic. Second-order transformation is sufficient in practice, but terms up to order ten were computed and timed to illustrate the fundamental improvement in automation and scalability over the commutator series approach.

**Spin relaxation theories**

Exponential integrals of the form discussed above also appear in the context of Bloch-Redfield-Wangsness spin relaxation theory[1,3], in which the relaxation superoperator $\mathbf{R}$ arising from the stochastic rotational modulation of spin interaction anisotropies has the following general form:

$$\mathbf{R} = -\sum_{kmpq} \mathbf{Q}_{km} \int_0^\infty G_{kmpq}(\tau) e^{-i\mathbf{H}_0 \tau} \mathbf{Q}_{pq}^\dagger e^{i\mathbf{H}_0 \tau} d\tau , \tag{19}$$

where $\mathbf{H}_0$ is the time-independent part of the spin Hamiltonian commutation superoperator, $\mathbf{Q}_{km}$ are the 25 irreducible spherical components of its anisotropic part $\mathbf{H}_1(t)$, defined so that[13]

$$\mathbf{H}_1(t) = \sum_{km} \mathfrak{D}_{km}^{(2)}(t) \mathbf{Q}_{km} . \tag{20}$$

The rotational correlation functions $G_{kmpq}(t)$ are defined as ensemble averages (denoted by angular brackets) of products of second-rank Wigner $\mathfrak{D}$ functions of molecular orientation:

$$G_{kmpq}(t) = \left\langle \mathfrak{D}_{km}^{(2)}(t) \mathfrak{D}_{pq}^{(2)*}(0) \right\rangle. \tag{21}$$

In systems undergoing stochastic motion these functions may be expressed as linear combinations of decaying exponentials[29,30]:

$$G(t) = \sum_n a_n e^{-\lambda_n t} \tag{22}$$

that are scalars and therefore commute with all matrices. With this observation in place, we can conclude that individual matrix-valued integrals in Equation (19) have the following general form:

$$\int_0^\infty e^{-i\mathbf{H}_0 t} \mathbf{Q} e^{i(\mathbf{H}_0 + i\lambda \mathbf{1})t} dt \tag{23}$$

These integrals used to be evaluated by diagonalizing $\mathbf{H}_0$ and computing the Fourier transform analytically[1,3,11,12] as shown in Equation (2). Diagonalization has cubic cost and eigenvector arrays of sparse spin Hamiltonians are full – this has limited the applicability of BRW theory to systems with fewer than about six spins. An improvement was suggested by Kuprov in 2011 – he pointed out that exponentiating $\mathbf{H}_0$ is cheaper than diagonalizing it and suggested a numerical quadrature route to the same integral[13] that extended the range of accessible matrix dimensions into hundreds of thousands[25]. In this communication we point out that Equation (23) is a case of an auxiliary exponential relationship:

$$\int_0^T e^{-i\mathbf{H}_0 t}\mathbf{Q}e^{i(\mathbf{H}_0+i\lambda\mathbf{1})t}dt = \mathbf{A}^\dagger\mathbf{B}$$

$$\begin{pmatrix} \mathbf{A} & \mathbf{B} \\ \mathbf{0} & \mathbf{C} \end{pmatrix} = \exp\left[\begin{pmatrix} i\mathbf{H}_0 & \mathbf{Q} \\ \mathbf{0} & i\mathbf{H}_0 - \lambda\mathbf{1} \end{pmatrix}T\right]$$

(24)

in which we made use of the fact that $\mathbf{H}_0$ is Hermitian and replaced the inverse of its exponential propagator $\mathbf{A}$ with a conjugate-transpose. The upper integration limit should be set according to the accuracy goal for the relaxation superoperator as suggested by Kuprov[13]:

$$T > \ln(1/\xi)\tau_C^{\max}$$

(25)

where $\tau_C^{\max}$ is the longest decay time present in the stochastic motion autocorrelation function and $\xi$ is the desired relative accuracy of the integral.

The improvement in the numerical efficiency of large-scale relaxation superoperator calculations produced by Equation (24) is very significant – it was this method that has enabled the first protein-scale NOESY simulations reported in our recent paper[25], where matrices of dimension far exceeding $10^5$ had to undergo relaxation theory processing. A scaling diagram, using a few standard systems included into the *Spinach* example set[14], is given in Figure 1B. Note that the comparison is given relative to the quadrature method[13] – it is of course utterly impossible to diagonalize the 800,000 by 800,000 matrix that presents itself during the NOESY calculation for ubiquitin[25].

The improvement in performance may be rationalized using matrix operation counting. The computational complexity of Kuprov's numerical quadrature method[13], measured by the number of matrix-matrix multiplications involved, increases *linearly* with the rotational correlation time because the upper limit $T$ of the integral in Equation (24) becomes larger. The auxiliary matrix method is faster for large spin systems in Figure 1B because the computational complexity of the scaling and squaring procedure involved in matrix exponentiation is *logarithmic* with respect to $T$. Based on our practical experience with restricted state spaces and modern computing hardware, we can reasonably state that Equation (24) permits accurate quantum mechanical spin relaxation theory treatment, including all cross-correlations and non-secular pathways, for liquid state NMR systems with up to about 2,000 spins[14] when it is combined with restricted state space methods.

**Optimal control theories**

The task of taking a quantum system from one state to another to a specified accuracy with minimal expenditure of time and energy, with the emphasis on the word *minimal*, is increasingly important in physics and engineering[8,31-33]. Optimal solutions are usually found numerically, by maximizing "fidelity" – the overlap between the final state of the system and the desired destination state[9]:

$$f = \langle \boldsymbol{\delta} | \boldsymbol{\rho}(T) \rangle = \langle \boldsymbol{\delta} | \exp_{(O)} \left[ -i \int_0^T \left( \mathbf{H}(t) + i\mathbf{R} \right) dt \right] | \boldsymbol{\rho}(0) \rangle \tag{26}$$

where angular bracket denotes column-wise vectorization of the corresponding density matrix, $\boldsymbol{\rho}(0)$ is the initial stare, $\boldsymbol{\delta}$ is the desired destination state, $\mathbf{H}(t)$ is the Hamiltonian commutation superoperator, $\mathbf{R}$ is the relaxation superoperator and $\exp_{(O)}$ denotes a time-ordered exponential. The Liouvillian superoperator of the spin system, defined as $\mathbf{L}(t) = \mathbf{H}(t) + i\mathbf{R}$, typically contains the "drift" part $\mathbf{L}_0$ that cannot be influenced and the "control" part that the measurement instrument can vary within certain limits:

$$\mathbf{L}(t) = \mathbf{L}_0 + \sum_k c^{(k)}(t) \mathbf{L}_k \tag{27}$$

where $\mathbf{L}_k$ are the control operators (*e.g.* radiofrequency and microwave Zeeman operators) and $c^{(k)}(t)$ are their time-dependent coefficients. The maxima of the fidelity functional $f$ with respect to the control sequences $c^{(k)}(t)$, subject to experimental constraints on feasible amplitudes and frequencies, are the subject of the mathematical branch of the optimal control theory[34,35].

The standard *modus operandi* in spin dynamics (known in general as the GRAPE algorithm[32]) is to discretize time and solve Equation (26) using thin slice propagators. On a finite grid of time points $0 \le t_n \le T$, the control sequences $c^{(k)}(t)$ become vectors:

$$c^{(k)}(t) = c_n^{(k)}, \quad t_{n-1} < t < t_n \tag{28}$$

and the task of minimizing the fidelity functional becomes a high-dimensional non-linear optimization problem. Numerical optimization is remarkably well researched[36] and, in this context, gradient descent, quasi-Newton and Newton-Raphson families of methods are generally appropriate. From the computational efficiency point of view, the central problem is therefore the calculation of first and second derivatives of the fidelity functional:

$$\frac{\partial f}{\partial c_n^{(i)}} = \langle \boldsymbol{\delta} | \mathbf{P}_N \cdots \mathbf{P}_{n+1} \frac{\partial \mathbf{P}_n}{\partial c_n^{(i)}} \mathbf{P}_{n-1} \cdots \mathbf{P}_1 | \boldsymbol{\rho}_0 \rangle$$

$$\frac{\partial^2 f}{\partial c_n^{(i)} \partial c_m^{(j)}} = \langle \boldsymbol{\delta} | \mathbf{P}_N \cdots \mathbf{P}_{n+1} \frac{\partial \mathbf{P}_n}{\partial c_n^{(i)}} \mathbf{P}_{n-1} \cdots \mathbf{P}_{m+1} \frac{\partial \mathbf{P}_m}{\partial c_m^{(j)}} \mathbf{P}_{m-1} \cdots \mathbf{P}_1 | \boldsymbol{\rho}_0 \rangle \tag{29}$$

where time slice propagators (assuming a fixed time grid step $\Delta t$) are defined as:

$$\mathbf{P}_n = \exp\left[ -i \left( \mathbf{L}_0 + \sum_k c_n^{(k)} \mathbf{L}_k \right) \Delta t \right] \tag{30}$$

The primary task therefore is to calculate first and second directional derivatives of the slice propagators. This problem is also very well studied[16] and the solution has already been mentioned in Equation (18). Restricting our attention to first and second derivatives specifically, we obtain the following auxiliary matrix relations:

$$\begin{pmatrix} \mathbf{P}_n & \dfrac{\partial \mathbf{P}_n}{\partial c_n^{(i)}} & \dfrac{1}{2}\dfrac{\partial^2 \mathbf{P}_n}{\partial c_n^{(i)} \partial c_n^{(j)}} \\ 0 & \mathbf{P}_n & \dfrac{\partial \mathbf{P}_n}{\partial c_n^{(j)}} \\ 0 & 0 & \mathbf{P}_n \end{pmatrix} = \exp\left[-i\begin{pmatrix} \mathbf{L}_0 + \sum_k c_n^{(k)}\mathbf{L}_k & \mathbf{L}_i & 0 \\ 0 & \mathbf{L}_0 + \sum_k c_n^{(k)}\mathbf{L}_k & \mathbf{L}_j \\ 0 & 0 & \mathbf{L}_0 + \sum_k c_n^{(k)}\mathbf{L}_k \end{pmatrix}\Delta t\right] \quad (31)$$

Because the number of control channels enumerated by $k$ is small, each individual propagator $\mathbf{P}_n$ in Equation (30) only depends on a small number of coefficients $c_n^{(k)}$. This means that the number of first and second propagator derivatives that must be computed to obtain the fidelity functional Hessian in Equation (29) is *linear* with respect to the number of time steps in the control sequence. This is unusual – the computational cost of the Hessian matrix is more commonly found to be quadratic in the number of parameters[36]. A cheap Hessian makes it possible to use Newton-Raphson type optimization algorithms that have better convergence properties than gradient descent and its variations[9,32]. An exploration of the convergence properties of the Newton-Raphson version of the GRAPE algorithm is outside the scope of this work and will be published separately.

**Radical pair dynamics**

Haberkorn[6] and Jones-Hore[5] models of radical pair recombination stipulate the following equations of motion for the spin density operator $\mathbf{\rho}(t)$:

$$\begin{aligned} \frac{d\mathbf{\rho}(t)}{dt} &= -i[\mathbf{H},\mathbf{\rho}(t)] - \frac{k_\mathrm{S}}{2}(\mathbf{\rho}(t)\mathbf{P}_\mathrm{S} + \mathbf{P}_\mathrm{S}\mathbf{\rho}(t)) - \frac{k_\mathrm{T}}{2}(\mathbf{\rho}(t)\mathbf{P}_\mathrm{T} + \mathbf{P}_\mathrm{T}\mathbf{\rho}(t)) + \mathbf{R}\mathbf{\rho}(t) \\ \frac{d\mathbf{\rho}(t)}{dt} &= -i[\mathbf{H},\mathbf{\rho}(t)] - (k_\mathrm{S} + k_\mathrm{T})\mathbf{\rho}(t) + k_\mathrm{S}\mathbf{P}_\mathrm{T}\mathbf{\rho}(t)\mathbf{P}_\mathrm{T} + k_\mathrm{T}\mathbf{P}_\mathrm{S}\mathbf{\rho}(t)\mathbf{P}_\mathrm{S} + \mathbf{R}\mathbf{\rho}(t) \end{aligned} \quad (32)$$

where $\mathbf{H}$ is the spin Hamiltonian, $\mathbf{R}$ is the spin relaxation superoperator, $\mathbf{P}_\mathrm{S,T}$ are two-electron singlet and triplet projection operators, $k_\mathrm{S,T}$ are singlet and triplet radical pair recombination rates and the assumption of first-order chemical kinetics unencumbered by spatial diffusion effects is made for the radical pair. Convenient solutions are possible in the adjoint representation (*aka* Liouville space[11]), where both models acquire the same general form

$$\frac{d}{dt}|\mathbf{\rho}(t)\rangle = -i\mathbf{L}|\mathbf{\rho}(t)\rangle \quad (33)$$

in which $|\boldsymbol{\rho}(t)\rangle$ denotes a column-wise vectorization of $\boldsymbol{\rho}(t)$ and

$$\begin{aligned} \mathbf{L}_H &= \mathbf{H}^- - \frac{i}{2}\left(k_S \mathbf{P}_S^+ + k_T \mathbf{P}_T^+\right) + i\mathbf{R} \\ \mathbf{L}_{JH} &= \mathbf{H}^- - i(k_S + k_T)\mathbf{1} + ik_S \mathbf{P}_T \otimes \mathbf{P}_T^T + ik_T \mathbf{P}_S \otimes \mathbf{P}_S^T + i\mathbf{R} \end{aligned} \quad (34)$$

for Haberkorn and Jones-Hore model respectively. In Equation (34), the upper T index indicates matrix transpose operation and $\mathbf{1}$ is a unit matrix of an appropriate dimension. Commutation and anti-commutation superoperators are defined as

$$\mathbf{O}^\pm = \mathbf{O} \otimes \mathbf{1} \pm \mathbf{1} \otimes \mathbf{O}^T \quad (35)$$

The target properties in radical pair simulations are singlet and triplet product yields $Y_{S,T}(t)$, defined as time integrals over the singlet and triplet populations:

$$Y_S(t) = k_S \int_0^t \langle \mathbf{P}_S | e^{-i\mathbf{L}t'} | \boldsymbol{\rho}_0 \rangle dt' \qquad Y_T(t) = k_T \int_0^t \langle \mathbf{P}_T | e^{-i\mathbf{L}t'} | \boldsymbol{\rho}_0 \rangle dt' \quad (36)$$

where $\boldsymbol{\rho}_0$ is the density matrix for the initial state. Evaluation of these integrals is the slowest stage of radical pair dynamics simulations because either a diagonalization[4] or a matrix inverse-times-vector operation[37] is normally involved. It is clear, however, that both may be avoided because the yields in Equation (36) are matrix elements of an exponential integral that may be computed using a special case of Equation (3) with $|\mathbf{0}\rangle$ denoting an all-zero vector of the same size as $|\boldsymbol{\rho}\rangle$:

$$Y_S(t) = k_S \int_0^t \langle \mathbf{P}_S | e^{-i\mathbf{L}t'} | \boldsymbol{\rho}_0 \rangle dt' = k_S \langle \mathbf{P}_S | \int_0^t e^{-i\mathbf{L}t'} dt' | \boldsymbol{\rho}_0 \rangle = k_S \left(\langle \mathbf{P}_S | \quad \langle \mathbf{0}|\right) \exp\left[\begin{pmatrix} \mathbf{0} & \mathbf{1} \\ \mathbf{0} & -i\mathbf{L} \end{pmatrix} t\right] \begin{pmatrix} |\mathbf{0}\rangle \\ |\boldsymbol{\rho}_0\rangle \end{pmatrix} \quad (37)$$

and similarly for the triplet yield. The importance of using Krylov propagation for $\exp(t\mathbf{A})\vec{v}$ type operations[38] should be stressed – it is not necessary to compute the entire matrix exponential when only one element is required by Equation (37).

Similar solutions are possible for the simplified kinetic models in which radical pair recombination is assumed to be a first-order process with a time-dependent recombination probability[4]:

$$Y_S(t) = \int_0^t \text{Tr}\left[e^{-i\mathbf{H}t'} \boldsymbol{\rho}_0 e^{i\mathbf{H}t'} \mathbf{P}_S\right] f(t') dt' \qquad Y_T(t) = \int_0^t \text{Tr}\left[e^{-i\mathbf{H}t'} \boldsymbol{\rho}_0 e^{i\mathbf{H}t'} \mathbf{P}_T\right] f(t') dt' \quad (38)$$

The recombination probability function $f(t)$ may be expanded as

$$f(t) = \sum_n k_n e^{-k_n t} \quad (39)$$

and the integrals in Equation (38) transformed into a form that permits the application of the auxiliary matrix exponential technique. For the singlet yield case:

$$\int_0^t \text{Tr}\left[e^{-i\mathbf{H}t'}\boldsymbol{\rho}_0 e^{i\mathbf{H}t'}\mathbf{P}_\text{S}\right]ke^{-kt'}dt' = k\text{Tr}\left[\int_0^t e^{-i\mathbf{H}t'}\boldsymbol{\rho}_0 e^{i\mathbf{H}t'}\mathbf{P}_\text{S}e^{-kt'}dt'\right] =$$
$$= k\text{Tr}\left[\mathbf{P}_\text{S}\int_0^t e^{-i\mathbf{H}t'}\boldsymbol{\rho}_0 e^{(i\mathbf{H}-k\mathbf{1})t'}dt'\right] \quad (40)$$

The integral under the trace is another instance of the problem already treated in Equation (24):

$$\int_0^t e^{-i\mathbf{H}t'}\boldsymbol{\rho}_0 e^{i(\mathbf{H}+ik\mathbf{1})t}dt = \mathbf{A}^\dagger \mathbf{B}$$
$$\begin{pmatrix} \mathbf{A} & \mathbf{B} \\ 0 & \mathbf{C} \end{pmatrix} = \exp\left[\begin{pmatrix} i\mathbf{H} & \boldsymbol{\rho}_0 \\ 0 & i\mathbf{H}-k\mathbf{1} \end{pmatrix}t\right] \quad (41)$$

The advantage of this method over the diagonalization technique is that matrix sparsity is preserved in the Hamiltonian exponentiation procedure[14], leading to large memory and CPU time savings[4]. Another practically relevant reminder is that Frobenius scalar product may be computed very efficiently, particularly for sparse matrices, by element-wise multiplication:

$$\langle \mathbf{A}|\mathbf{B}\rangle = \text{Tr}\left[\mathbf{A}^\dagger \mathbf{B}\right] = \text{Tot}\left[\mathbf{A}^* \odot \mathbf{B}\right] \quad (42)$$

where $\mathbf{A}^\dagger$ indicates the conjugate-transpose of $\mathbf{A}$, $\mathbf{A}^*$ indicates the element-wise complex conjugate of $\mathbf{A}$, $\odot$ denotes element-wise matrix product and $\text{Tot}$ stands for the total sum.

**Conclusions**

The primary advantage of auxiliary matrix techniques for the theories described in this paper is logistical simplicity – assembling a block matrix, exponentiating it and extracting a block from the result is easier and neater than summing a Magnus expansion or a commutator series.

Another major advantage is numerical efficiency, particularly when the matrices involved in the simulation are sparse. Spin dynamics is special in this regard because spin Hamiltonians in the Pauli basis are guaranteed to be very sparse[20]. This sparsity is destroyed by common matrix operations – diagonalization, inversion and singular value decomposition would in general produce full arrays – but preserved by exponentiation when the matrix is appropriately scaled. Recognising this fact and recasting physical theories in a way that specifically prefers matrix exponentiation to diagonalization leads to significant improvements in numerical efficiency and scaling.

**Acknowledgements**

We are grateful to Sophie Schirmer and Luke Edwards for useful discussions. This work was made possible by EPSRC through a grant to IK group (EP/H003789/1) and a CDT studentship to DLG.

**Figure captions**

**Figure 1**  Scaling diagrams for the augmented matrix routes through the theories described in the main text. (**A**) wall clock time taken by the $^{14}$N rotating frame transformation to the specified order, applied to the spin system of methylaziridine. The molecular geometry, chemical shielding tensors, *J*-couplings and nuclear quadrupolar interaction tensors were computed using GIAO DFT M06/cc-pVDZ[39-41] method in Gaussian09[42]. More details on the spin system and the relaxation theory problem in question are available from our recent paper on the subject[28]. The calculation uses the restricted state space approximation[43,44] with IK-0(4) basis set[25], the reduced state space dimension is 9,889. (**B**) Wall clock time (in seconds) comparison between the matrix-valued numerical quadrature method for the calculation of the integral in Equation (23) and the auxiliary matrix method presented in this paper. The spin systems indicated in the figure come from the *Spinach* library[14] example set.

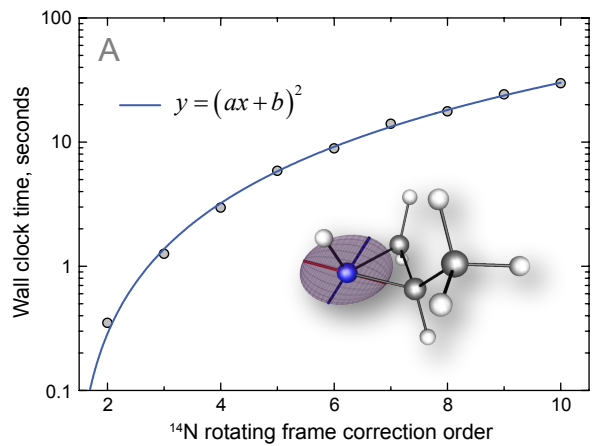 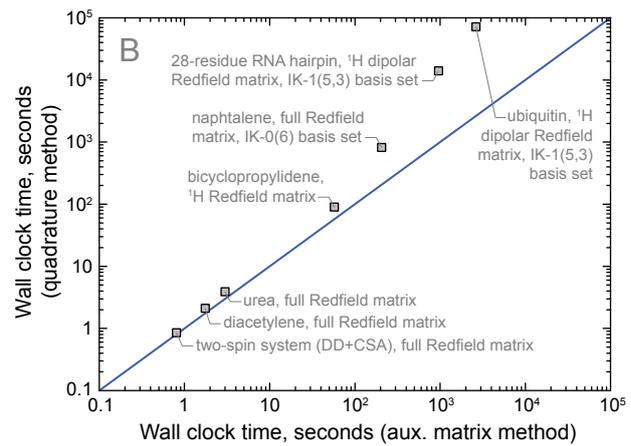